\documentclass[amssymb,amsmath,superscriptaddress,footinbib,notitlepage]{revtex4-1}
\usepackage{natbib,graphicx,subfigure,subeqnarray,fancyhdr,amsmath,multirow,color, mathrsfs}
\usepackage{graphicx}
\usepackage{epstopdf, epsfig}
\usepackage{color}
\usepackage{amsmath,amssymb,amsfonts}

\usepackage[colorlinks=true, citecolor=blue, linkcolor=blue,urlcolor=blue ]{hyperref}

\newcommand{\pd}[2]{\frac{\partial #1}{\partial #2}} 
 
\newcommand{\bn}{\boldsymbol{n}}
\newcommand{\bu}{\boldsymbol{u}}
\newcommand{\be}{\boldsymbol{e}}
\newcommand{\bx}{\boldsymbol{x}}

\newcommand\Da{\mbox{\textit{Da}}}

\begin{document}

\title{Phoretic flow induced by asymmetric confinement}

\author{Maciej Lisicki}
\email{m.lisicki@damtp.cam.ac.uk}
\affiliation{Department of Applied Mathematics and Theoretical Physics, University of Cambridge, UK.}
\affiliation{Institute of Theoretical Physics, Faculty of Physics, University of Warsaw, Poland.}
\author{S\'{e}bastien Michelin}
\affiliation{LadHyX - D\'{e}partement de M\'{e}canique, Ecole Polytechnique - CNRS, France.}
\author{Eric Lauga}
\affiliation{Department of Applied Mathematics and Theoretical Physics, University of Cambridge, UK.}

\begin{abstract}
Internal phoretic flows due to the interactions of solid boundaries with local chemical gradients 
may be created using chemical patterning. Alternatively, we demonstrate here that internal flows might also be induced by geometric asymmetries of chemically-homogeneous surfaces. We characterise the circulatory flow created in a cavity enclosed between two eccentric cylindrical walls of uniform chemical activity. Local gradients of the diffusing solute induce a slip flow along the surface of the cylinders, leading to a circulatory bulk flow pattern which can be solved analytically in the diffusive limit. The flow strength can be controlled by adjusting the relative positions of the cylinders and an optimal configuration is identified. These results provide a model system for tunable phoretic pumps. 
\end{abstract}

\date{\today}

\maketitle

\section{Introduction}

Microscopic flow generation lies at the root of many technological and biological processes. With the advent of microfluidic manipulation techniques, fluids flowing in narrow channels may be used for medical diagnostics, biological sensing, and nano-scale chemical synthesis, and  it thus becomes essential to  induce and control flow within a confined geometry \cite{squires2005,whitesides2006}. Classically, this has been  achieved by imposing a global external pressure gradient in the system or by surface forcing, either using electrokinetic effects or by generating stresses at boundaries to locally drive the flow.  The latter mechanism is widely exploited by biological systems. Coordinated beating of biological flagella leads to self-propulsion of micro-swimmers \cite{brennen1977,lauga2009}.  
A complementary process of microscale pumping involves directional flow driven by boundaries, a  mechanism used, for example, in embryonic development \citep{hirokawa2009,johnson2016}, in the mammalian reproductive tract \cite{halbert1976}, or for the transport of mucus in human lungs \cite{sleigh1988}. Recently, artificial devices have been designed in which surface flow is generated by flapping flagella of entrapped bacteria \cite{gao2015}.

In a coarse-grained view, all these surface-driven processes share the common feature that the bulk flow is generated by imposing the motion of the fluid on the confining boundary. Classically, phoretic motion is induced by externally-imposed gradients, such as electric field, temperature, or concentration \cite{anderson1984}. In artificial biomimetic systems, this may also be achieved using phoretic effects where the interactions between a surface and an external field gradient create a local flow in the boundary's close vicinity \cite{anderson1989,derjaguin1947} which then may be used for self-propulsion {\cite{golestanian2007,sharifimood2013,michelin2014,howse2007}}, migration of particles in externally imposed chemical gradients, or flow generation \cite{gao2012}. Local directional flow may be achieved using  chemical micropatterning~\cite{yadav2015}, anisotropic wettability of the channel surface \cite{wang2015}  or by spatially varying surface charges of the walls \cite{stroock2003}. 

An alternative  method to locally induce phoretic flow with chemical patterning is to exploit geometrical asymmetries in the channel walls \cite{michelin2015,yang2015}.  In this paper we further explore this concept by presenting a model system in which the internal flow may be fully controlled by geometry. 
We consider a cavity consisting of two eccentric cylinders and calculate  the flow generated by the chemical activity of its surfaces. Assuming uniform chemical activity of the walls, where the solute is consumed (or released) at a fixed  rate, a fully analytical solution for the concentration and flow fields may be obtained. We determine the flux in the resulting flow as a function of the cavity's geometry, characterised by the size ratio of the cylinders and the eccentricity in their position. An optimal configuration can be identified in which the fluid motion is maximum. An important aspect of the considered system is that the flow rate may be controlled solely by displacing the inner cylinder from a symmetric configuration, where there is no motion by symmetry, to the most eccentric, which maximises the volumetric flux of the fluid within. This system represents thus a tunable internal phoretic pump.

The paper is organised as follows. The general framework of phoretic motion and its solution using bi-polar coordinates are first presented in \S\ref{sec_general}. In \S\ref{sec_results}, we analyse the resulting flow field and optimise the geometry to achieve the maximal possible rate of the circulating flow. Finally, \S\ref{sec_conclusions} summarizes the main findings and presents some perspectives. 

%
%

\section{Phoretic flow generation between two eccentric cylinders}\label{sec_general}

\subsection{Continuum phoretic framework}
We follow the continuum framework of phoretic motion {\cite{golestanian2007,julicher2009,sabass2012}} and consider the two-dimensional flow generated in a cavity contained between two eccentric and chemically-homogeneous cylinders of circular cross-section $\mathcal{S}_1$ and $\mathcal{S}_2$, filled by a fluid of dynamic viscosity $\eta$ and density $\rho$, containing a solute species of local concentration $C$ and diffusivity $\kappa$. Assuming that the chemical properties of the outer wall maintain a uniform concentration $C_0$ on $\mathcal{S}_1$, we focus in the following on the relative concentration $c=C-C_0$. The inner wall's chemical activity $\mathcal{A}$ quantifies the fixed rate of solute release ($\mathcal{A}>0$) or absorption ($\mathcal{A}<0$) at the surface:
\begin{equation}\label{fixedflux}
\kappa \bn\cdot\nabla c =-\mathcal{A} \textrm{      on   } \mathcal{S}_2,
\end{equation}
with $\bn$ being a unit vector normal to the surface. Because of the short-range interaction of solute molecules with the cavity's boundary, local concentration gradients result in pressure forces imbalances and the fluid being set into motion. Assuming that the interaction layer thickness is small compared to the cavity's dimensions, the classical slip-velocity formulation may be used~{{\cite{anderson1989,golestanian2007}}}, and local solute gradients result in a net slip velocity 
\begin{equation}\label{slip_def}
\bu =\mathcal{M}(\boldsymbol{1}-\bn\bn)\cdot\nabla c\textrm{      on   } \mathcal{S}_j,\qquad j=1\textrm{  or  }2,
\end{equation}
that drives a flow within the cavity. Here, $\mathcal{M}$ is the local phoretic mobility of the cylinders' surface. Note that, since the concentration is uniform on the outer boundary $\tau=\tau_1$, Eq.~\eqref{slip_def} results in the classical no-slip boundary condition there ($\bu=\mathbf{0}$). When the P\'eclet number $\mathrm{Pe}=\mathcal{UR}/\kappa$ is small enough (e.g. small solute molecules), the solute dynamics is purely diffusive and is thus governed by the Laplace's equation
\begin{equation}\label{laplacec}
\nabla^2 c= 0,
\end{equation}
in the fluid domain. Here, $\mathcal{R}$ denotes the radius of $\mathcal{S}_1$ chosen as characteristic length in this work, and $\mathcal{U}=|\mathcal{AM}|/\kappa$ is the characteristic phoretic velocity generated along $\mathcal{S}_2$.  Provided that inertial effects are negligible (i.e. $\mathrm{Re}=\rho\, \mathcal{U} \mathcal{R}/\eta\ll 1$), the flow and pressure fields satisfy the incompressible Stokes equations
\begin{align}\label{Stokes}
\nabla\cdot\bu = 0,\quad \eta\nabla^2 \bu =\nabla p.
\end{align}
The diffusive Laplace's problem for the solute thus effectively decouples from the hydrodynamic problem and may be solved independently. Its solution for the concentration $c$ can be used to compute the slip flow along $\mathcal{S}_2$, Eq.~\eqref{slip_def}, which then determines the flow field within the cavity by solving Eq.~\eqref{Stokes}. 

The problem is non-dimensionalised as follows: with $\mathcal{R}$ and $\mathcal{U}$ chosen as characteristic length and velocity respectively, the characteristic concentration scale is set by $|\mathcal{A}|\mathcal{R}/\kappa$ and the characteristic pressure reads $\eta |\mathcal{A}\mathcal{M}|/\mathcal{R} \kappa$. 

\begin{figure}
\begin{center}
\includegraphics[width=0.4\columnwidth]{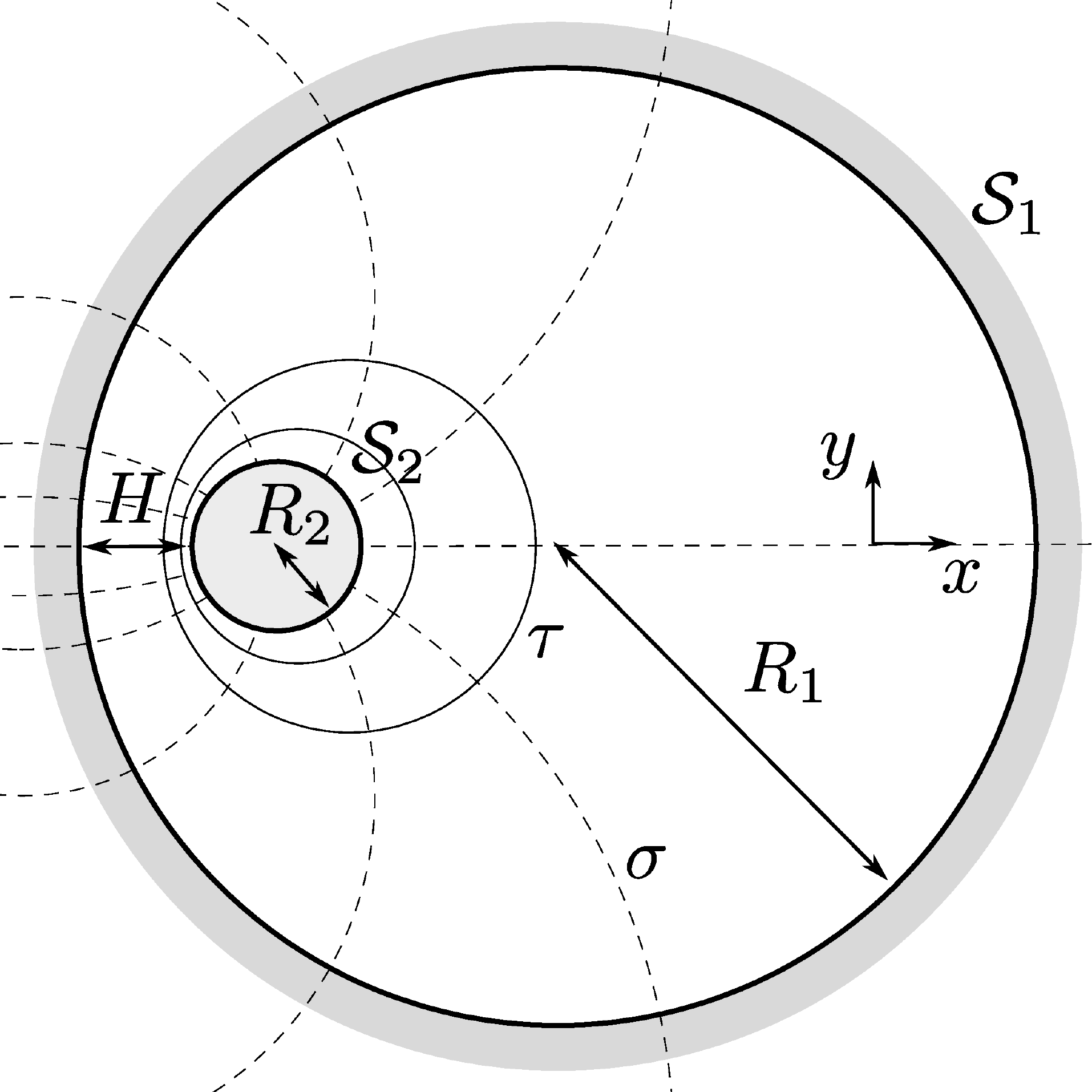}
\caption{Notation for the annular cavity consisting of two non co-axial cylinders.  Bi-polar  coordinates are represented by surfaces of constant $\tau$ (solid) and $\sigma$ (dashed). The surfaces $\tau=\tau_1$ and $\tau=\tau_2$  correspond to the boundaries of the outer and inner cylinder, respectively. }\label{fig:schema}
\end{center}
\end{figure}

\subsection{Formulation of the non-dimensional problem}\label{sec_formulation}
The two-dimensional cavity $\Omega_f$ is enclosed between two eccentric cylindrical surfaces $\mathcal{S}_1$ and $\mathcal{S}_2$ of respective non-dimensional radii $R_1=1$ and $R_2$. The centres of the cylinders are offset by a distance $d$ (Fig.~\ref{fig:schema}). The diffusion problem for solute concentration reads
\begin{align}
\nabla^2c=0&\textrm{      in  } \Omega_f,\\ \label{flux_bc}
 c=0 \textrm{      on   } \mathcal{S}_1,&\qquad\bn\cdot\nabla c=-A \textrm{      on   } \mathcal{S}_2, 
\end{align}
{with the dimensionless activity $A=\mathcal{A}/|\mathcal{A}|$}, and the associated hydrodynamic flow problem may now be formulated as
\begin{align}
\nabla^2 \bu = \nabla p &\textrm{      in  }\Omega_f, \\ \label{slip_bc}
\bu = \boldsymbol{0} \textrm{      on   } \mathcal{S}_1, \qquad &
\bu  = M (\boldsymbol{1}-\bn\bn)\cdot\nabla c \textrm{      on   } \mathcal{S}_2, 
\end{align}
with $M=\mathcal{M}/|\mathcal{M}|=\pm 1$ the dimensionless mobility. 

\subsection{Computation of the flow field}\label{sec_flow}

\subsubsection{Bi-polar coordinates}

It is natural in this problem to introduce the bipolar cylindrical coordinate system, $-\infty<\tau<\infty$, $-\pi \leq \sigma < \pi$, related to the Cartesian coordinates $(x,y)$  by
\begin{equation}\label{xy}
x = \frac{a\sin\sigma}{\cosh\tau - \cos\sigma} , \quad y = \frac{a\sinh\tau}{\cosh\tau - \cos\sigma}.
\end{equation}	
with $x$ being horizontal in Fig.~\ref{fig:schema}, and the scale factors given by
\begin{equation}
h_\sigma = h_\tau = h = \frac{a}{\cosh\tau - \cos\sigma}= \frac{a}{\sinh\tau}\left(1 + 2\sum_{n=1}^{\infty} e^{-n\tau}\cos n\sigma\right).\label{eq:scale}
\end{equation}
{The last equality follows from representation of the scale factor $h$ in terms of a Fourier series in $\sigma$. Since $h$ is an even function of $\sigma$, the series has only cosine terms.} The unit vectors $\be_\tau$ and $\be_\sigma$, normal to the isolines of $\tau$ and $\sigma$, respectively, are given by $\pd{\bx}{\tau} = h_\tau \be_\tau$ and $\pd{\bx}{\sigma}=h_\sigma \be_\sigma$, and form an orthonormal basis in 2D.

The isolines  $\tau=\tau_0$ are circles of radius $a/|\sinh\tau_0|$ centred on the $y$-axis at $y=a\coth\tau_0$. In the following, we choose $\tau_{1,2}>0$, i.e. both circular boundaries lay in the upper half-plane; their centres   are separated by a distance $d$. The parameters of the system $(d,R_1,R_2)$ satisfy thus
\begin{equation}\label{geometricparams}
R_{1,2} = \frac{a}{\sinh\tau_{1,2}},\quad d=a(\coth\tau_1-\coth\tau_2).
\end{equation}
The closest distance between the circles, the gap width $H$, is given by $H=R_1-R_2-d$. 
\subsubsection{Solute diffusion}
The Laplace's equation for the concentration field in two dimensions becomes
\begin{equation}\label{Laplace_bipolar}
\frac{1}{h^2} \left(\pd{^2 c}{\tau^2} + \pd{^2 c}{\sigma^2}\right) = 0.
\end{equation}
The Dirichlet boundary condition on the  outer cylinder is $c(\tau_1,\sigma)=0$. Noting that $\bn(\tau)=-\be_\tau$ for $\tau=\tau_2$, the flux boundary condition, Eq.~\eqref{flux_bc}, follows  as
\begin{equation}
\be_\tau\cdot\nabla c = \frac{1}{h} \pd{c}{\tau} = A.
\end{equation}
Equation~\eqref{Laplace_bipolar} is separable, and using Eq.~\eqref{eq:scale}, its solution satisfying these boundary conditions is
\begin{align}\label{conc_analytical}
c(\tau,\sigma)=AR_2(\tau-\tau_1) + 2AR_2 \sum_{n=1}^{\infty} \frac{e^{-n\tau_2}}{n}\frac{\sinh [n(\tau-\tau_1)]}{\cosh [n(\tau_1-\tau_2)]} \cos n \sigma.
\end{align}

\subsubsection{Stokes flow}
The Stokes flow problem may be conveniently formulated in bi-polar cylindrical coordinates using the stream function $\psi(\tau,\sigma)$, such that
\begin{equation}\label{velocity_psi}
u_\tau = -\frac{1}{h} \pd{\psi}{\sigma}, \qquad u_\sigma =  \frac{1}{h}\pd{\psi}{\tau}\cdot
\end{equation}
The stream function is related to the flow vorticity by $\omega= - \nabla^2 \psi$. Since the vorticity $\boldsymbol{\omega}=\nabla\times\bu$ is harmonic, $\nabla^2 \boldsymbol{\omega}=0$, we conclude that  $\psi$ satisfies the biharmonic equation
\begin{equation}\label{biharmonic}
\nabla^4 \psi= \nabla^2 \nabla^2 \psi = 0,
\end{equation}
with the Laplacian given in bipolar cylindrical coordinates in Eq.~\eqref{Laplace_bipolar}. As shown by Jeffery \cite{jeffery1921}, it is most convenient to consider the function $\Psi = \psi/h$ for which Eq.~\eqref{biharmonic} becomes a linear equation with constant coefficients
\begin{equation}
\left( \pd{^4}{\tau^4} + 2\pd{^4}{\tau^2\partial\sigma^2} +\pd{^4}{\sigma^4} - 2\pd{^2}{\tau^2} + 2\pd{^2}{\sigma^2} + 1 \right)\Psi = 0,
\end{equation}
with  solution also  given by Jeffery \cite{jeffery1922}.  The boundary conditions on the cylinders are then written in terms of $\Psi=\psi/h$ using Eq.~\eqref{velocity_psi} and simplify, noting that because of symmetry and boundary conditions, the axis of symmetry and both cylinders are on the same streamline. Therefore, $\Psi(\tau_{1,2},\sigma)=0$. The remaining velocity boundary conditions read
\begin{align}
\pd{\Psi}{\tau} &=  0,\ \textrm{      on   } \tau=\tau_1, \\
\pd{\Psi}{\tau} &=  u_\sigma(\tau_2,\sigma)\ \textrm{      on   } \tau=\tau_2.
\end{align}
The slip velocity on the surface of the inner cylinder is obtained from Eq.~\eqref{slip_def} using the solution of the solute concentration problem \eqref{conc_analytical} as 
\begin{align}\label{slipflow}
u_\sigma(\tau_2,\sigma) &= \frac{ M A R_2}{a} \Bigl[(d_2-2d_1\cosh\tau_2)\sin\sigma + \sum_{n=2}^{\infty} (d_{n+1} + d_{n-1} - 2 d_n \cosh\tau_2) \sin n\sigma \Bigr],
\end{align}
with $d_n = e^{-n\tau_2}\tanh n(\tau_2-\tau_1)$. The resulting stream function reads
\begin{align}\label{solution_psi}
\psi &=  h\Psi =  \frac{MAR_2}{\cosh\tau - \cos\sigma}\Biggl[(\alpha_1\cosh 2\tau + \beta_1 + \gamma_1\sinh2\tau + \delta_1 \tau)\sin\sigma\Biggr. + \\ \nonumber
&+\Biggl.\sum_{n=2}^{\infty}\Bigl(\alpha_n \cosh(n+1)\tau + \beta_n\cosh(n-1)\tau + \gamma_n\sinh(n+1)\tau + \delta_n\sinh(n-1)\tau\Bigr)\sin n\sigma  \Biggr],
\end{align}
with  coefficients listed in Appendix \ref{app_coeff}. This explicit form of the stream function may now be used to characterise the flow.

%
%

\section{Results and discussion}\label{sec_results}
\begin{figure}
\begin{center}
\includegraphics[width=0.5\columnwidth]{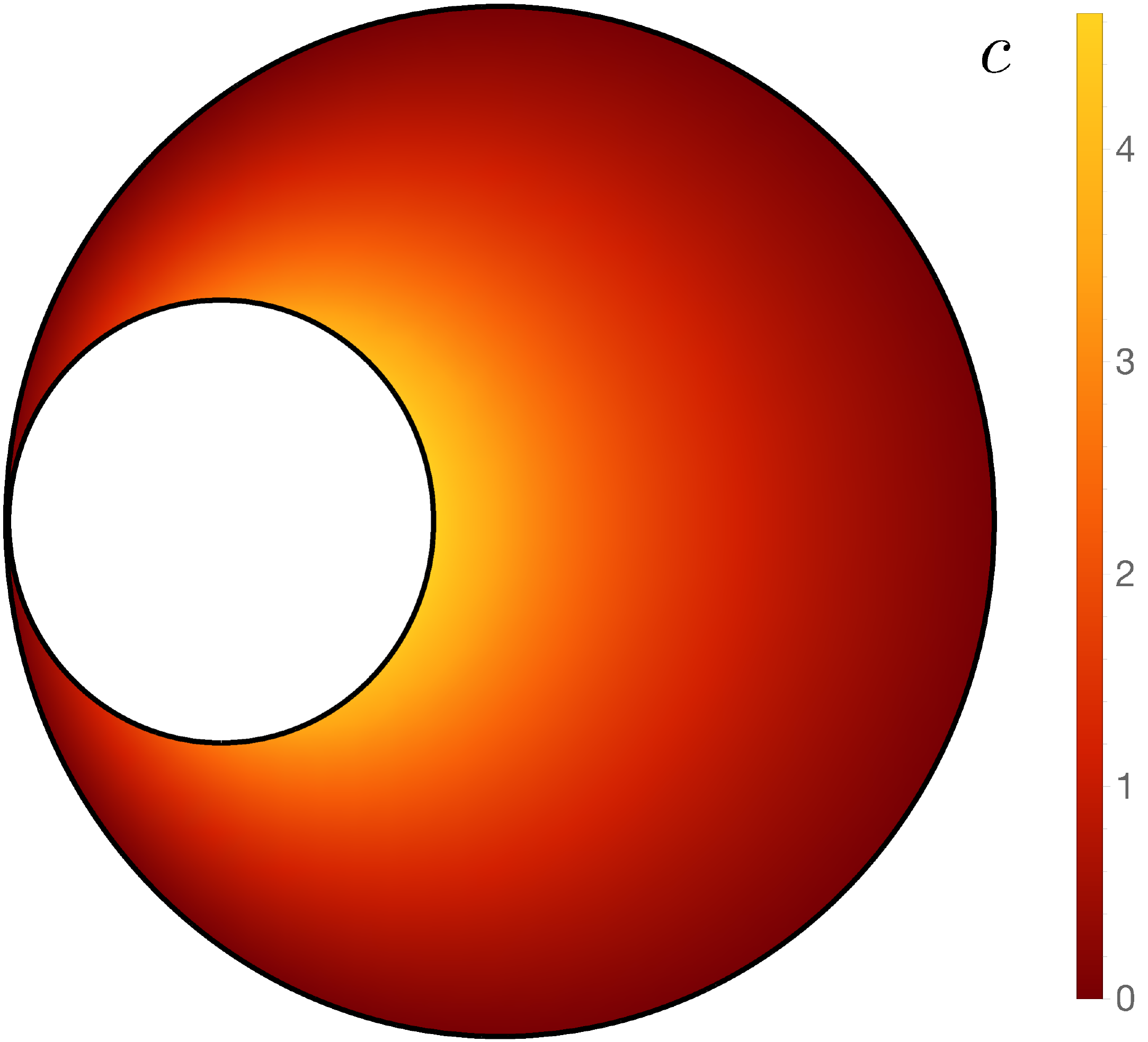}\includegraphics[width=0.5\columnwidth]{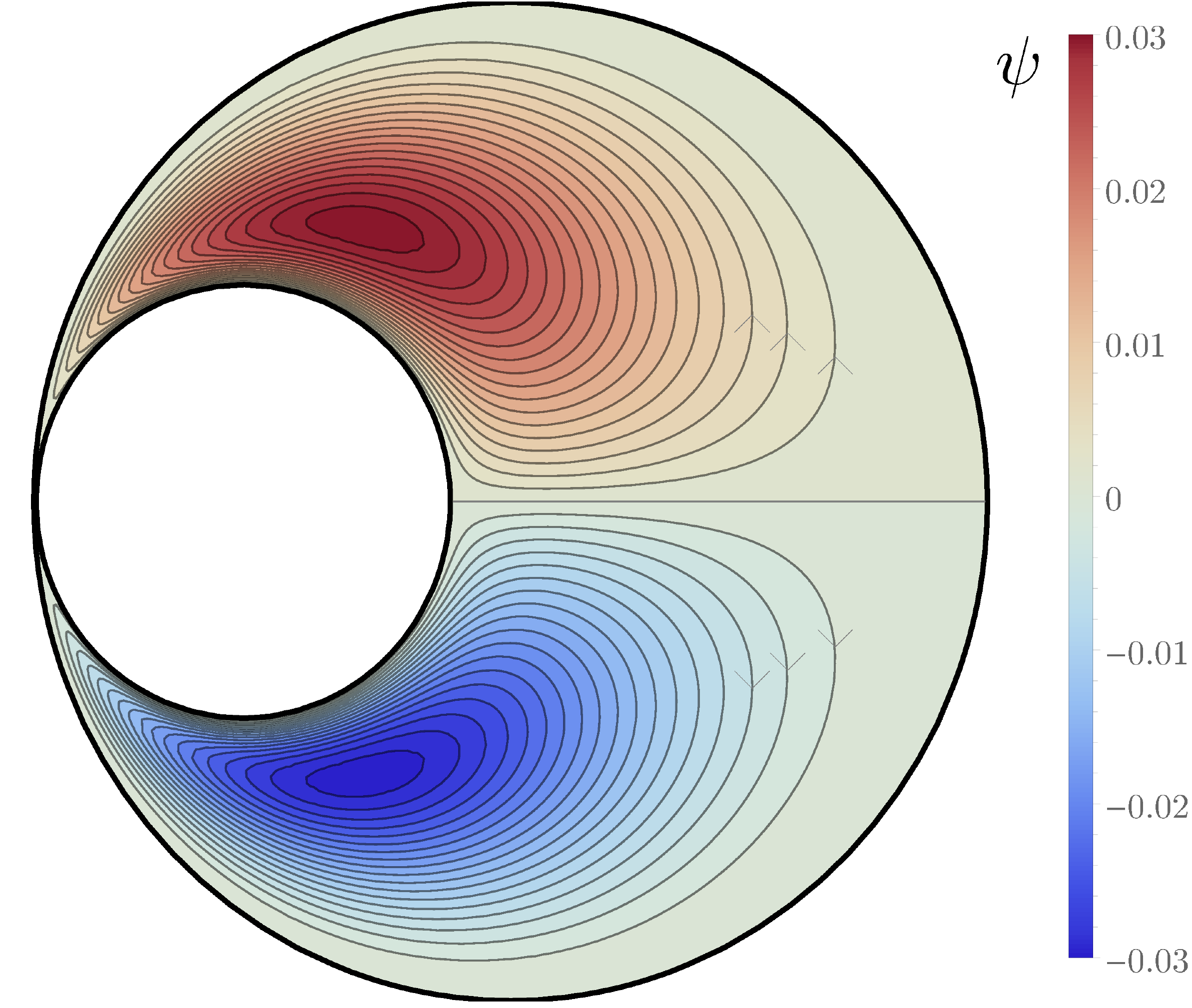}
\caption{Distribution of  solute concentration $c$ (left) and streamfunction $\psi$ (right)  within the cavity for $R_2/R_1 = 0.43 ,H/H_\mathrm{max}=0.01$ and ${A}=1$ (solute released by the inner cylinder) and $M=1$ (the flow direction is reversed when $AM=-1$). The isolines of $\psi$ are the   streamlines.}\label{fig:conc}
\end{center}
\end{figure}

In non-dimensional form, the flow resulting from the chemical activity of the bounding surfaces now solely depends on two geometric parameters, namely the relative radii of the cylinders $R_2/R_1$ and the asymmetry parameter $d/R_1$ controlling the eccentricity. The latter can be alternatively described by the relative gap width, $0\leq H/R_1\leq 1-R_2/R_1$. {The maximal gap width thus reads $H_\mathrm{max}=R_1-R_2$.}

{For the needs of demonstration, in all of the following figures we truncate the series \eqref{conc_analytical} and \eqref{solution_psi} at a finite, sufficiently large $n$.} A typical solute concentration profile in the eccentric annular cavity is depicted in Fig.~\ref{fig:conc}.  For $A>0$ (resp. $A<0$), the requirement of a constant normal gradient at the surface of the inner cylinder, Eq. \eqref{flux_bc}, results in the concentration of solute being highest (resp. lowest) in the central area of the cavity, and decreasing (resp. increasing) along the surface when moving towards the most confined region between the cylinders. In this narrower gap, tangential concentration gradients are lower due to the shorter distances allowed for   diffusive transport. As a result, tangential gradients of concentration and slip flow arise on the surface of the inner cylinder. For $AM>0$, they are oriented from the narrower gap to the central part of the cavity. Two stagnation points are found on the axis of symmetry and the slip flow is maximal on the lateral sides of the inner cylinder.

This boundary forcing generates two  counter-rotating cells within the cavity, with the flow speed being maximal at the surface of the inner (driving) cylinder. This is most easily demonstrated in Fig.~\ref{fig:conc}, where the streamlines are plotted. Note that the flow, and thus its direction, depends on the sign of chemical activity and phoretic mobility of the boundary: if $AM>0$, the  vorticity  is positive in the upper cell and negative in the lower cell.

An appealing property of the present setup is the ability  to control the magnitude of the flow by tuning the geometry. In the context of constructing an optimal flow-inducing  device in such a geometry, it is important to determine which configuration (size ratio and gap width) maximises the strength of flow. For a perfectly centred inner cylinder, the concentration distribution is isotropic within the cavity, leading to no flow forcing and no fluid motion. Eccentricity therefore plays a key role in driving the flow within the cavity, and most asymmetric configurations are expected to stir the fluid most efficiently.

\begin{figure}
\begin{center}
\includegraphics[width=0.6\columnwidth]{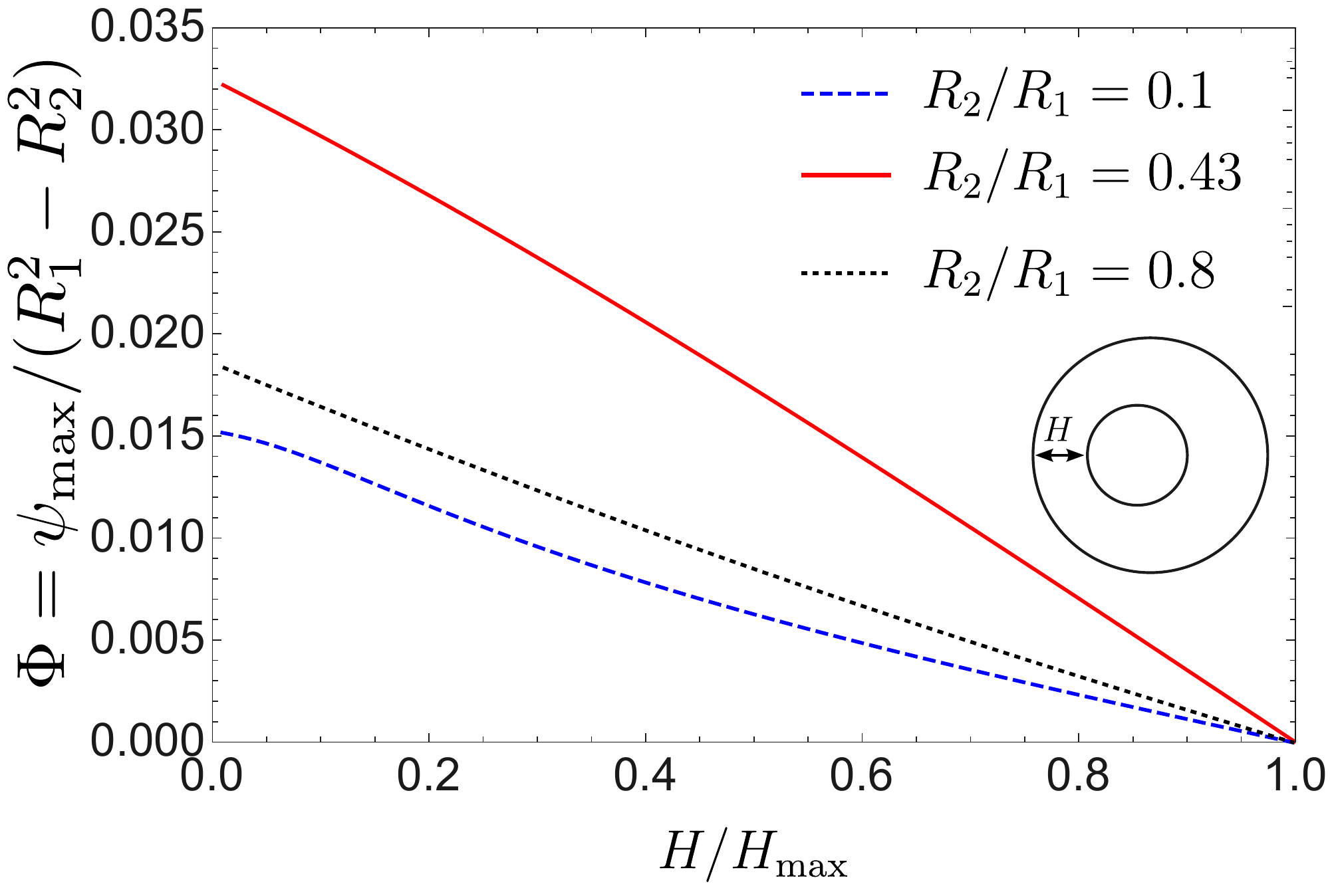}\\
\includegraphics[width=0.6\columnwidth]{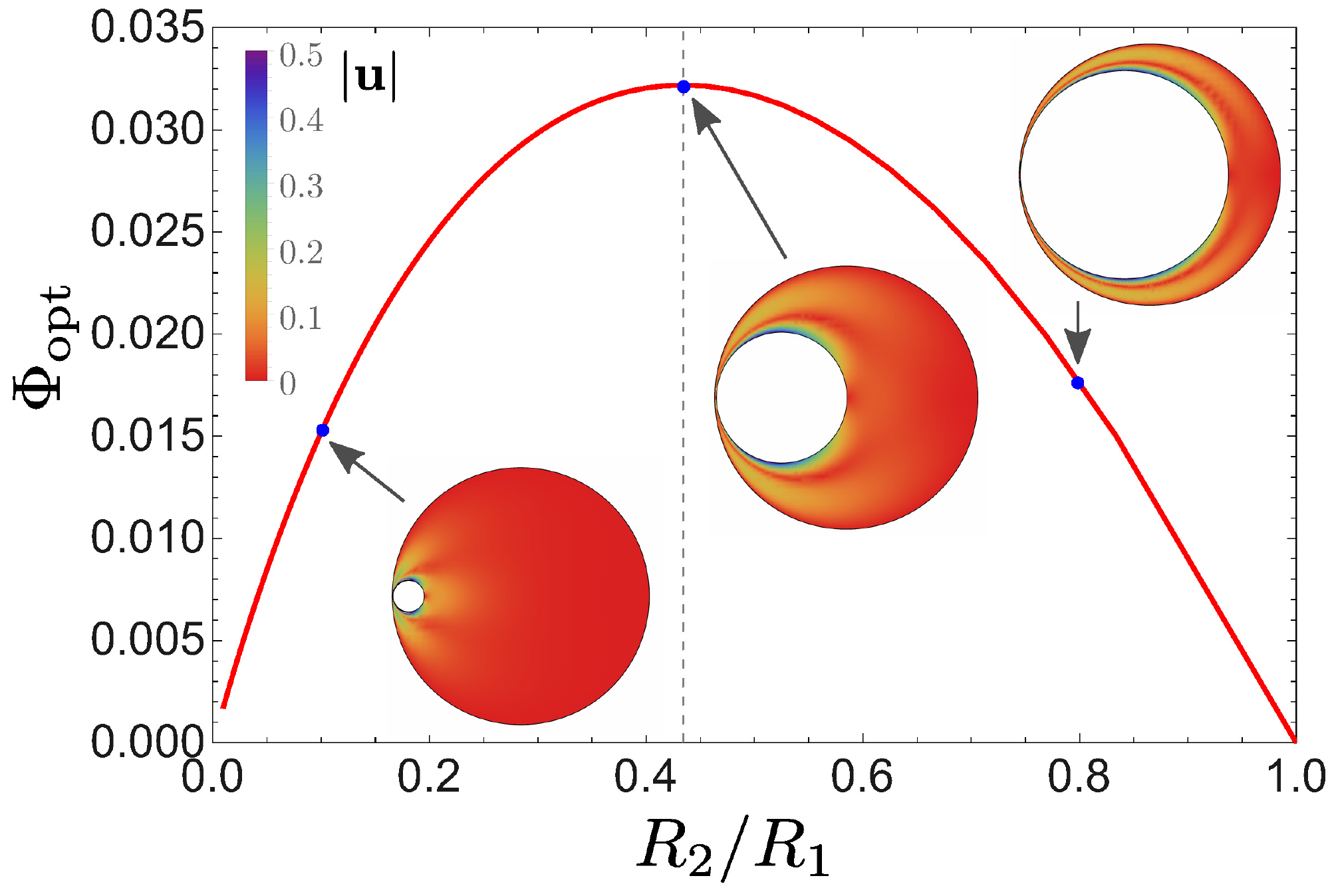}
\caption{(Top) Evolution of the scaled total flux in a circulation cell, $\Phi=\psi_\mathrm{max}/(R_1^2-R_2^2)$,  with the relative gap width $H/H_\textrm{max}$ for $R_2/R_1=0.1$, 0.43 and 0.8. (Bottom) Evolution of the optimal scaled flux $\Phi_\textrm{opt}$ (obtained for touching cylinders) with $R_2/R_1$. The magnitude of the velocity field is also shown for three cases (same as above) showing the competing effects of phoretic driving and wall hindrance on the flow.}
\label{fig_psi_max}
\label{fig_psi_max_h}
\end{center}
\end{figure}  

  In order to quantify that intuition and the strength of flow generated by the chemical activity on the surface of the inner cylinder, we calculate the total flux of the circulating fluid. Since the flux across a curve between two points $A$ and $B$ is  the difference of the stream functions $\psi(B)-\psi(A)$, and given that on the surfaces of the cylinders $\psi=0$, the total flux in the circulation cell is given by the maximal value of the stream function $\psi_\mathrm{max}$ in the fluid. In the following, we focus on the volume flux per unit area of the cavity, $\Phi=\psi_\textrm{max}/(R_1^2-R_2^2)$, as a measure of flow motion. An alternative interpretation of $1/\Phi$ also provides an estimate of the characteristic stirring time for the fluid, namely a weighted average of the period of fluid particles motion along the streamlines.
  
  For a given size ratio of the cylinders, we find an approximately linear increase of this flux with increasing asymmetry (decreasing the gap size $H$), with a maximum attained in the limit $H\rightarrow 0$ regardless of the size ratio (see Fig.~\ref{fig_psi_max_h}, top). This is consistent with the observation that auto-phoretic particles tend to be strongly repelled from a neighbouring wall, as recently investigated by Uspal {\it et al.} \cite{uspal2015} for a spherical active particle close to a planar wall. {In this limit, we look for the optimal size ratio which maximises the flow rate.}
  
 The value of the optimal scaled flux $\Phi_\textrm{opt}$ depends on the relative sizes of the cylinders. For large size ratios the motion of the fluid is hindered by the no-slip boundary condition on the outer boundary and leads to small flow. When the inner cylinder gets small, strong gradients induce  large slip velocities; however, these are capable of moving only the fluid in their immediate vicinity, resulting in most of the fluid remaining at rest. This is confirmed by Fig.~\ref{fig_psi_max} (bottom) where we plot the dependence with the size ratio of the maximum scaled volumetric flux $\Phi$ (obtained for touching cylinders). Clearly, for size ratios 0 and 1 the flow ceases, and a maximum is present at $R_2/R_1\approx 0.43$. 

\begin{figure}
\begin{center}
\includegraphics[width=0.6\columnwidth]{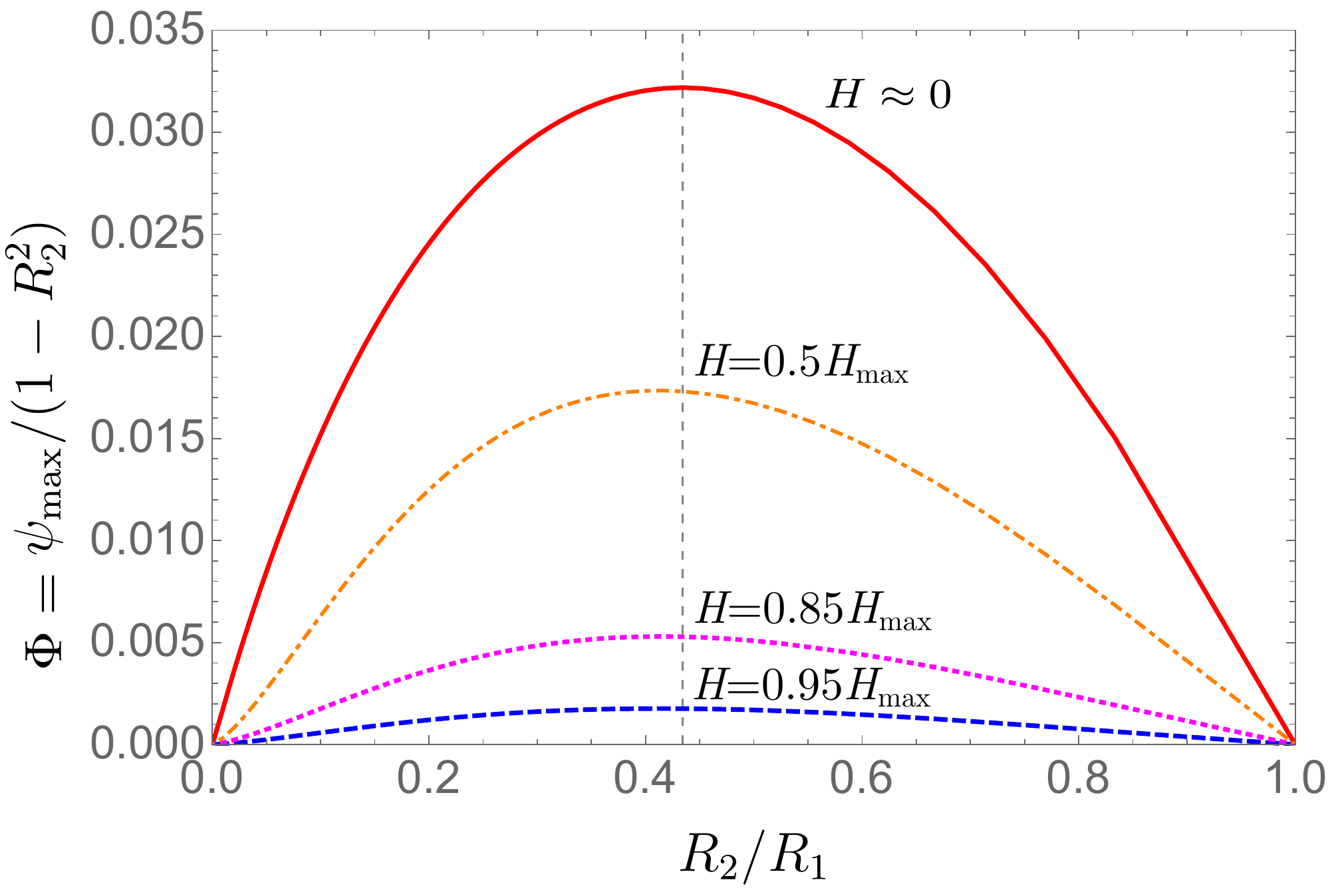}
\caption{{The dependence of the total flux $\Phi$ on the size ratio shows that the optimal size ratio $R_2/R_1 \approx 0.43$ is universal and approximately independent on the gap width  $H$ between the cylinders.}}
\label{fig_H}
\end{center}
\end{figure}

{Interestingly, our numerical results show the optimal size ratio which maximises the flow rate seems not to depend on the asymmetry parameter $H/H_\mathrm{max}$, and therefore is a universal feature of the system, as shown in Fig. \ref{fig_H}. This suggests that this ratio may be determined likewise in the case when $d\approx 0$, at only a slight asymmetry, and by taking only the $n=1$ term in Eq. \eqref{solution_psi}, as this is enough to grasp the character of the flow in this far-field limit. Numerical solution yields the optimal size ratio $R_2/R_1\approx 0.41$, that is within $5\%$ from the value determined at small gap using the full solution.}

  As an extension to the flow generation problem discussed above, we note that the fixed-flux condition \eqref{fixedflux} may be generalised to a one-step chemical reaction by assuming the activity to be proportional to the local solute concentration, $\mathcal{A}=-\mathcal{K}C$, with a constant reaction rate $\mathcal{K}$ \cite{cordova2008}. 
  The relative importance of the reaction rate and diffusive transport of solute is then quantified by the Damk\"ohler number $\Da = \mathcal{KR}/\kappa$. For $\Da\ll 1$ the diffusion of the solute is fast enough to ensure a homogeneous consumption of solute on the boundaries, while the limit $\Da\gg 1$ corresponds to slow diffusion being unable to compensate for the fast local reactive effects. As for the propulsion problem, the effect of an increasing Damk\"ohler number on our setup is a limitation of the inner wall's chemical activity by the depletion in solute resulting from the slow diffusion. As a result, the concentration gradients induced in the system are reduced, and therefore the resulting slip velocities are generally lower. The main conclusions of the previous analysis, namely the existence of an optimal configuration for touching cylinders and intermediate size ratio, remain valid nonetheless (not shown).

%
%
\section{Conclusions}\label{sec_conclusions}

Recent years have brought significant attention to auto-phoresis as a promising mechanism for microscale fluid manipulation and transport. The manufacturing of such miniature devices poses physical challenges which require a deeper understanding of the physical mechanisms underlying the phoretic  generation of viscous flows in confined geometries. Our paper provides an example of a system in which the flow can be fully controlled by adjusting its geometric configuration and a first attempt at optimising internal flows driven by phoretic effects. 

Specifically we have analysed the model system of a two-dimensional cavity between two eccentric cylinders. Bulk flow circulation can be induced within the cavity from the chemical activity of the walls (i.e. the release/absorption of solute) and their phoretic mobility (i.e. their interaction with local solute content). Neglecting solute advection, the problem has an analytical solution which was presented here in terms of the Stokes flow stream function. By analysing the explicit formulae, an optimal configuration was identified for the system, quantified by the maximal flow rate within the fluid volume, in terms of the position and size of the inner cylinder with respect to the outer one. Replacing  the fixed-flux boundary condition by a simple chemical reaction leads to a general decrease of the efficiency of the device but similar overall conclusion. 

Our results identify an optimal configuration, namely a maximum eccentricity achieved for two touching cylinders. In that case, the narrowest gap is effectively closed and the geometry of the system resembles that of a circular cavity with an inner protrusion. Optimal flow circulation then results from the concentration gradients enhanced by the local higher curvature of the boundary. This effect of curvature on phoretic flow enhancement is consistent with recent work on phoretic propulsion~\cite{michelin2015b}. The limit of touching cylinders deserves however more scrutiny; when $H$ becomes smaller than the typical interaction-layer thickness $\lambda$, the slip-velocity formulation of phoretic flows, Eq.~\eqref{slip_def} breaks down, and the above analysis should be replaced by a full description of the solute-wall interactions within the boundary layer.

\begin{acknowledgements}
This work was funded in part by a David Crighton Fellowship at the University of Cambridge (ML), a Mobility Plus Fellowship from the Polish Ministry of Science and Higher Education (ML), the EU through a Marie-Curie CIG grant (EL) and the French Ministry of Defense DGA (SM).
\end{acknowledgements}

\appendix

\section{Coefficients of the stream function}\label{app_coeff}

The exact solution for the stream function \eqref{solution_psi} involves the following coefficients which depend solely on $\tau_{1,2}$:
\begin{align}
\alpha_1 &= X\left\{2\Delta\tau\cosh 2\tau_1 - \sinh 2\tau_1 + \sinh 2\tau_2 \right\}, \\ \nonumber
\beta_1 &= X\left\{2\tau_2 - 2\tau_1 \cosh 2\Delta\tau + \sinh 2\Delta\tau \right\}, \\ \nonumber
\gamma_1 &= X\left\{-2\Delta\tau\sinh 2\tau_1 + \cosh 2\tau_1 - \cosh 2\tau_2 \right\},\\ \nonumber
\delta_1 &= 4 X \sinh^2\Delta\tau, \\ \nonumber
\alpha_n &= { Y_n\left\{n \sinh[2\tau_1 + (n-1)\tau_2]- \sinh[2 n \tau_1 + (1-n)\tau_2] - (n-1)\sinh (n+1)\tau_2  \right\},}\\ \nonumber
\beta_n &= { Y_n\left\{-(1+n)\sinh (n-1)\tau_2 - n \sinh[2\tau_1 - (n+1)\tau_2] + \sinh[2n\tau_1 - (n+1)\tau_2]  \right\},}\\ \nonumber
\gamma_n &= { Y_n\left\{-n \cosh[2\tau_1 + (n-1)\tau_2] + \cosh[2 n \tau_1 + (1-n)\tau_2] + (n-1)\cosh (n+1)\tau_2  \right\},}\\ \nonumber
\delta_n &= { Y_n\left\{(1+n)\cosh (n-1)\tau_2 - n \cosh[2\tau_1 - (n+1)\tau_2] - \cosh[2n\tau_1 - (n+1)\tau_2]  \right\}.}
 \end{align}
where $\Delta\tau=\tau_1-\tau_2$ and
\begin{align}
X &={\frac{d_2-2d_1\cosh\tau_2}{8\sinh^2{\Delta\tau} (1-\Delta\tau\coth\Delta\tau)}}, \\ \nonumber
Y_n &={\frac{1}{2} \frac{d_{n+1}+d_{n-1}-2 d_n \cosh\tau_2}{n^2-1-n^2\cosh 2\Delta\tau+\cosh 2n\Delta\tau}}.
\end{align}

\end{document}